\documentclass[aip,apl,10pt,twocolumn,superscriptaddress,reprint]{revtex4-1}

\usepackage{amsmath}
\usepackage{graphicx}
\usepackage{epstopdf}

\begin{document}

\title{Fast measurement of carbon nanotube resonator amplitude with a heterojunction bipolar transistor}
\date{\today}
\author{Kyle Willick}
\affiliation{Institute for Quantum Computing, University of Waterloo, Waterloo, Ontario, Canada}
\affiliation{Waterloo Institute for Nanotechnology, University of Waterloo, Waterloo, Ontario, Canada}
\affiliation{Department of Physics and Astronomy, University of Waterloo, Waterloo, Ontario, Canada}
\author{Xiaowu (Shirley) Tang}
\affiliation{Waterloo Institute for Nanotechnology, University of Waterloo, Waterloo, Ontario, Canada}
\affiliation{Department of Chemistry, University of Waterloo, Waterloo, Ontario, Canada}
\author{Jonathan Baugh}
\email[Contact: ]{baugh@uwaterloo.ca}
\affiliation{Institute for Quantum Computing, University of Waterloo, Waterloo, Ontario, Canada}
\affiliation{Waterloo Institute for Nanotechnology, University of Waterloo, Waterloo, Ontario, Canada}
\affiliation{Department of Chemistry, University of Waterloo, Waterloo, Ontario, Canada}

\begin{abstract}
Carbon nanotube (CNT) electromechanical resonators have demonstrated unprecedented sensitivities for detecting small masses and forces. The detection speed in a cryogenic setup is usually limited by the CNT contact resistance and parasitic capacitance. We report the use of a heterojunction bipolar transistor (HBT) amplifying circuit near the device to measure the mechanical amplitude at microsecond timescales. A Coulomb rectification scheme, in which the probe signal is at much lower frequency than the mechanical drive signal, allows investigation of the strongly non-linear regime. The behaviour of transients in both the linear and non-linear regimes is observed and modeled by including Duffing and non-linear damping terms in a harmonic oscillator equation. We show that the non-linear regime can result in faster mechanical response times, on the order of $10\;\mu\text{s}$ for the device and circuit presented, potentially enabling the magnetic moments of single molecules to be measured within their spin relaxation and dephasing timescales. 
\end{abstract}

\maketitle

Nanoscale mechanical resonators, such as suspended carbon nanotubes (CNTs), have demonstrated state-of-the-art sensitivity in mass \cite{Chiu2008,Chaste2012} and force \cite{Moser2013} detection due to the small CNT diameter, high Young's modulus, and the ability to achieve resonator quality factors (Q) up to 5 million \cite{Moser2015}. This extraordinary sensitivity has been used to detect masses down to the single proton level \cite{Chaste2012}. For more easily detected forces, one would like to harness this sensitivity to obtain fast measurements on timescales that compete with some dynamics of interest. For example, detection of a single molecule magnet grafted to a CNT resonator has been demonstrated using anisotropy-induced torsion \cite{Ganzhorn2013a}, and we calculated in a previous paper \cite{Willick2014} that with a suitable magnetic field gradient, mechanical frequency shifts on the order of 10 kHz could result from magnetic transitions of a single molecule. Single-shot observations of this magnetic switching would be of interest for investigating molecular nanomagnets individually (not in bulk), especially if the readout timescale can be comparable to the spin dephasing and relaxation times. Certain magnetic molecules of interest in quantum information processing have been found to dephase and relax on timescales of microseconds and milliseconds, respectively, at low temperatures \cite{Ardavan2007,Zadrozny2015}. An obstacle to fast measurements with CNTs, however, is the contact resistance which is typically in the range $\sim 25\; \text{k}\Omega$ for low-bandgap CNTs. When the device is separated from the first amplification stage by a transmission cable with stray capacitance $\sim 100\; \text{pF/m}$, the resulting RC time constant prevents fast observations. To overcome this problem, the electrical signal from the mechanical resonator must be amplified near the device. 

Silicon-Germanium heterojunction bipolar transistors (HBT) have been used for near-device cryogenic amplification to measure single electron transistors \cite{Curry2015}. HBTs have operating frequencies up to  hundreds of MHz, low power dissipation, and a gain $\sim 100$. In this letter, an HBT amplifier circuit in proximity to a CNT mechanical resonator is used to minimize readout capacitance and thus increase the measurement bandwidth. Fast sensing of CNT mechanical resonance has previously been demonstrated using a HEMT voltage amplifier and two-source mixing readout \cite{Meerwaldt2013b,Schneider2014}. Here we show that similar results can be obtained with a non-mixing readout technique using an HBT current amplifier. The readout is based on a modified Coulomb rectification technique, and has the advantage that relatively strong measurement signals can be used without affecting the mechanical motion, as the measurement signal is at a far lower frequency. We use this configuration to study the transient response of the CNT resonator to pulsed driving in both the linear and strongly non-linear regimes, finding that the latter regime can give a faster response in force sensing applications.  

\begin{figure}
\includegraphics[width=\linewidth]{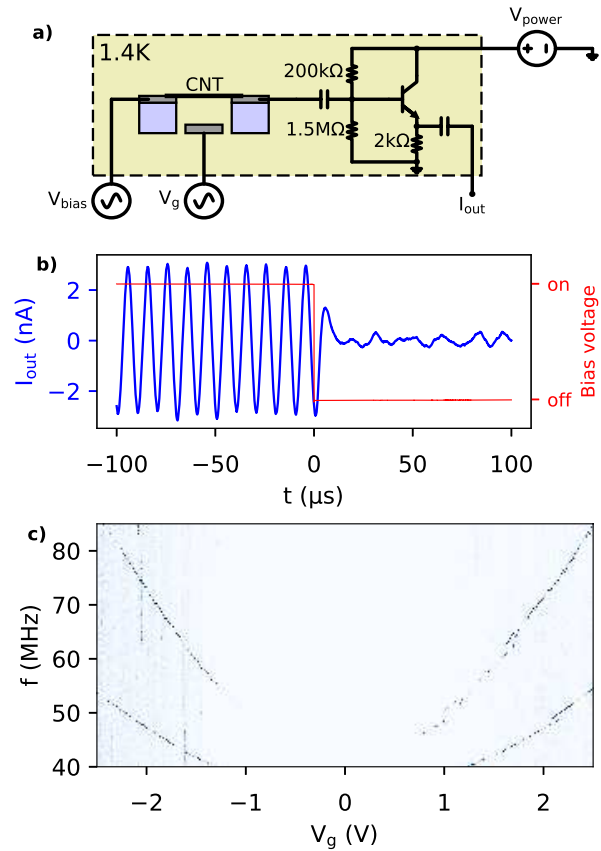}
\caption{\textbf{a}. The measurement setup used for fast readout of CNT mechanical motion. The CNT resonator is connected to a nearby common-collector HBT amplifier circuit, all at a temperature of 1.4 K. The capacitors are both 22 nF, and the HBT is an Infineon BFP842ESD powered with a DC voltage $V_{power} = 1.15 \text{V}$. \textbf{b}. Current response measured at $I_{out}$ when a 100 kHz, 1 mV signal on $V_{bias}$ is switched off (at $t=0$). The electrical response shows a $5 \mu\text{s}$ delay and a $\lesssim 10 \mu\text{s}$ ringdown, which is shorter than the typical timescale of CNT mechanical response. The current fluctuations observed for $t > 0$ indicate the noise floor of the measurement setup, approximately 0.15 nA. \textbf{c}. Mixing current (arbitrary units) observed for the CNT resonator in a previous measurement without the HBT circuit, using frequency-modulated mixing and an AC driving voltage of $500 \mu\text{V}$. The lower frequency mode is more weakly coupled to the gate, and is likely an in-plane mode. The higher frequency mode, which we focus on in this paper, is strongly coupled to the gate and is likely out-of-plane.}
\label{fig1}
\end{figure}

When used in a common-collector amplifier configuration, an HBT realizes a high-impedance AC current amplifier. Figure \ref{fig1}a shows the schematic of the measurement circuit we employed. A DC resistor network is used to switch the HBT into the amplification regime via a supplied voltage, $V_{power} = 1.15\;\text{V}$. The resistors and capacitors were selected to pass frequencies $\sim$100 kHz while minimizing power dissipation. The estimated power dissipation of this circuit is $120\;\mu\text{W}$, which is dominated by the DC current through the $2\;\text{k}\Omega$ emitter resistor. All measurements were carried out in a pumped $^4$He cryostat with the sample in vacuum at a temperature of 1.4 K.  

To realize fast measurements, the electrical ringdown time of the measurement circuit should be short compared to the mechanical response timescale. Figure \ref{fig1}b shows the averaged electrical current, $I_{out}$, of the HBT circuit when switching off an AC bias voltage $\text{V}_{bias}$ with no mechanical driving (i.e. $\text{V}_{g,ac}=0$). The applied $\text{V}_{bias}$ is a 100 kHz sine wave with 1 mV amplitude, which is switched off at $t = 0$. There is a $5\;\mu\text{s}$ delay between the applied pulse (either switching on or off) and the response observed through the measurement setup. The electrical ringdown has a time constant $\tau \lesssim 10\;\mu\text{s}$. This ringdown time could be further reduced by adjusting the HBT circuit parameters. 

The suspended CNT device was fabricated starting from an undoped silicon wafer with a 300 nm thick thermal silicon oxide layer. A 10/20 nm thick W/Pt layer was sputtered onto the wafer to serve as the contact metal. After photolithographic patterning, the Pt, W, and 240 nm of SiO$_2$ was etched to define a $2\; \mu\text{m}$ wide trench between contacts. A 10/30 nm thick Ti/Pt gate was patterned and evaporated into the bottom of the trench to act as a local gate electrode. In the final step, the CNT was grown across the trench using chemical vapour deposition at $850^{\circ}\text{C}$ with Fe-Co-Mo catalyst islands patterned on the W/Pt contacts near the trench edge.

The CNT device was characterized in a previous cool-down at 1.4 K using a frequency-modulation technique \cite{Gouttenoire2010} without the HBT circuit present. Figure \ref{fig1}c shows the mixing current measured with strong driving ($V_{g,ac} = 500\;\mu\text{V}$), as a function of the DC component of the gate voltage $\text{V}_{g,dc}$, where two primary modes were observed. The lower frequency mode is only weakly coupled to the gate, and is not observed at lower driving power. It is expected that this mode is in-plane with the gate, while the higher frequency mode is out-of-plane. The mode separation at low $V_g$ can be a result of built in tension anisotropy due to contacts\cite{Eichler2012}. In this paper we focus on the strongly gate-coupled out-of-plane mode.

To detect the CNT mechanical motion,  we use a modified Coulomb rectification readout scheme \cite{Huttel2009}. The mechanical driving signal, $V_{g,ac}$, is applied to the gate along with a DC gate voltage to electrically tune the device, $V_{g,dc}$. An AC voltage is applied to bias the device, as in figure \ref{fig1}b. When the bias frequency is much lower than the mechanical driving, the observed current at the bias frequency will be time-averaged over the mechanical oscillations, so that 
\begin{equation} \label{eq:rect}
I \approx I' + \left(\frac{x_0}{2}\frac{dC_g}{dx} \frac{V_{g,dc}}{C_g}\right)^2 \frac{\partial^2 I'}{\partial V_g^2}
\end{equation}
where $I$ is the measured current, $I'$ is the current in absence of mechanical driving, $C_g$ is the gate-CNT capacitance, $x$ is the distance between the CNT and gate, and $x_0$ is the amplitude of motion of the CNT. The dependence of $I$ on $\partial^2 I'/\partial V_g^2$ constrains this readout to measurements at non-linear conductance features such as Coulomb peaks. The Coulomb rectification readout method requires no component of the measurement signal (bias) to be near the resonance frequency, as is required for mixing techniques. This allows for large measurement voltages without inducing mechanical driving. This is especially advantageous when measuring strongly nonlinear resonators, where large drive powers near the resonance frequency result in broad resonant linewidths.

\begin{figure}
\includegraphics[width=\linewidth]{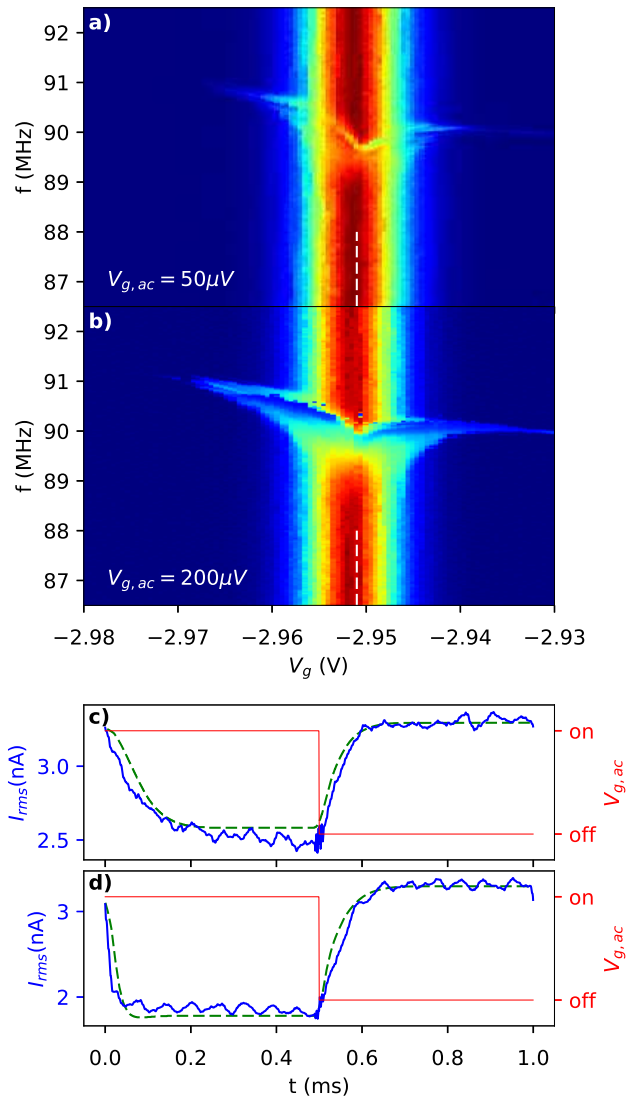}
\caption{\textbf{a,b} Current through the CNT mechanical resonator versus DC gate voltage (horizontal axis) and mechanical drive frequency (vertical axis). The color scale represents $I_{out}$ from 0 (dark blue) to 3.5 nA (red). The current is measured via lock-in amplifier with an applied bias voltage of 1 mV at 100 kHz. The mechanical drive is an AC gate voltage of $50\;\mu\text{V}$ and $200\;\mu\text{V}$, respectively. Nonlinear behaviour is clearly observed at the higher driving power. White dashed line indicates $V_g$ used in c,d. \textbf{c} Transient current response due to pulsed driving, related to the CNT mechanical amplitude by equation \ref{eq:rect}. This data was measured on the Coulomb peak in (a) and with $50\;\mu\text{V}$ drive voltage. The red line shows when $V_{g,ac}$ is applied, the blue line shows the measured current envelope, and the green dashed line shows a fit using equation \ref{eq:ho} with $\alpha = 5 \times 10^9\:\text{kg}\;\text{m}^{-2}\:\text{s}^{-2}, \gamma = 10^{-17}\:\text{kg}\;\text{s}^{-1}, \eta = 22\;\text{kg}\;\text{m}^{-2}\:\text{s}^{-1}$. \textbf{d} Transient response under the same conditions except that $V_{g,ac} = 200\;\mu\text{V}$. The green line fit uses the same $\alpha, \gamma, \eta$ values as (c).}
\label{fig2}
\end{figure}

\begin{figure}
\includegraphics[width=\linewidth]{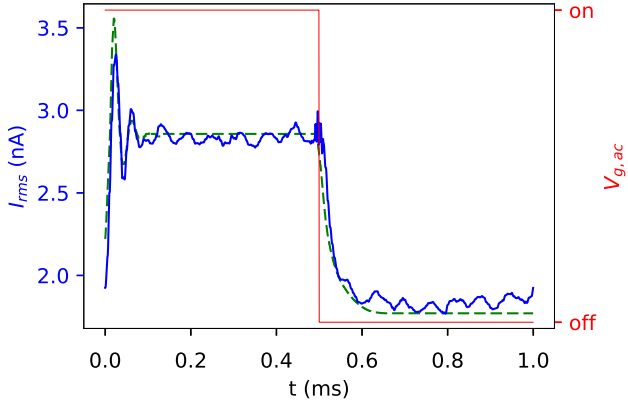}
\caption{Transient current response showing mechanical overshoot at the turn-on of a $50\;\mu\text{V}$ drive voltage. The red line shows when $V_{g,ac}$ is applied, the blue line shows the measured current envelope, and the green dashed line shows a fit using equation \ref{eq:ho} with $\alpha = -6 \times 10^{10}\:\text{kg}\;\text{m}^{-2}\:\text{s}^{-2}$, $\gamma = 3 \times 10^{-17}\:\text{kg}\;\text{s}^{-1}$, and $\eta = 100\;\text{kg}\;\text{m}^{-2}\:\text{s}^{-1}$.}
\label{fig3}
\end{figure}

The results in figures \ref{fig2} and \ref{fig3} use the rectification scheme, with a 1 mV, 100 kHz AC bias voltage. For the measurements in figure \ref{fig2}(a) and (b), $I_{out}$ from the HBT circuit is measured through a current-voltage preamplifier and input to a lock-in amplifier. For time-domain measurements, $I_{out}$ is amplified and recorded on an oscilloscope. The time-domain data is post-processed to extract the root mean square envelope function (see supplementary material).

Figure \ref{fig2}(a) and (b) shows the mechanical resonance measured across a Coulomb peak (bright vertical region), with driving voltages of $V_{g,ac} = 50\;\mu\text{V}$ and $200\;\mu\text{V}$, respectively. For low driving power, on the conductance peak we observe a small decrease in both the mechanical frequency and Q factor, as seen in similar devices and described by the coupling of single electron tunneling to mechanical modes \cite{Lassagne2009,Huttel2010,Steele2009a}. The stronger driving power in figure \ref{fig2}b causes nonlinear effects to become visible, and a switch from spring softening to spring hardening nonlinearities is seen, as was observed in reference \cite{Steele2009a}. Nonlinear behaviour is typical in CNT resonators due to their small dimensions and the electro-mechanical coupling of single electron tunneling.

Figure \ref{fig2} (c) and (d) show time-domain measurements of time-averaged mechanical motion in which transient decays are observed upon switching the drive on or off. Both measurements were performed at the maximum of the conductance peak shown in (a,b), and at the drive frequency 89.7 MHz, which is the mechanical resonance frequency in the linear (low drive power) regime. $V_{g,ac}$ is switched on/off at 1 kHz with $50\%$ duty cycle. In figure \ref{fig2}c, $V_{g,ac} = 50\;\mu\text{V}$ and the resonator shows transient mechanical ringdown expected for a linear harmonic oscillator driven on resonance. The signal decay time constant gives $Q \approx 2 \times 10^4$. At the higher drive power $V_{g,ac} = 200\; \mu\text{V}$ in figure \ref{fig2}(d), a similar transient response is seen for turn-off of the drive, but the turn-on has a more rapid response due to resonator non-linearity. 

To fit this behaviour, the oscillator was modelled using a Duffing non-linearity and non-linear damping in the harmonic oscillator equation \cite{Eichler2011a}
\begin{equation}\label{eq:ho}
m \frac{d^2x}{dt^2} + (k + \alpha x^2) x + (\gamma + \eta x^2) \frac{dx}{dt} = F_{drive} \cos(\omega t)
\end{equation}
where $m$ is the effective mass of the CNT mode, $k = m \omega_0^2$ is the linear spring constant, $\gamma$ is the linear damping coefficient, $\alpha$ is the Duffing parameter, $\eta$ is the non-linear damping coefficient, and $F_{drive} = (\partial C_g/\partial x) V_{g,dc} V_{g,ac}$  is the electrostatic driving force. The parameters in (\ref{eq:ho}) are initially estimated from the device geometry and adjusted to fit the transient response (see supplementary material). The time evolution is solved numerically, and converted to a current using off-resonance transport data and equation (\ref{eq:rect}). Importantly, the fits in figure \ref{fig2} (c) and (d) use the same set of parameters, with only the value of $V_{g,ac}$ being different.

Measurements at a second Coulomb peak revealed further non-linearity in the transient response. Figure \ref{fig3} shows transient current measured at the edge of a conductance peak at $V_{g,dc} = -3.03\;V$ (see supplementary material), using low drive power $V_{g,ac} = 50\; \mu\text{V}$. Sitting at the edge of the Coulomb peak maintains a large $\partial^2 I'/\partial V_g^2$ for the rectification signal, but changes the non-linear parameters, in particular changing the sign of $\alpha$. In addition, this Coulomb peak had a higher conductance than that measured in figure \ref{fig2}, which is expected to give a stronger non-linearity. In this case, the transient signal shows an overshoot \cite{Monroe2012} of the steady state amplitude during turn-on, which is also captured by equation \ref{eq:ho}. In this regime, the non-linear amplitude response pulls the effective resonance frequency away from the drive frequency, creating a damped beating similar to a linear oscillator with off-resonant driving. 

\begin{figure}
\includegraphics[width=\linewidth]{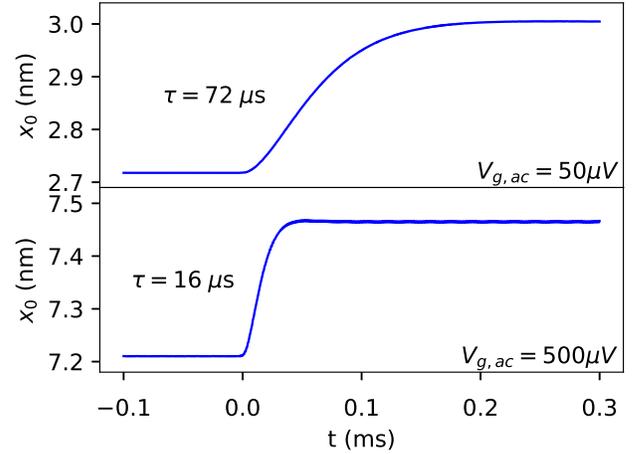}
\caption{Simulated transient response to a resonance frequency shift of 10 kHz at $t = 0$. The resonator is modelled with $\alpha = 5 \times 10^{9}\:\text{kg}\;\text{m}^{-2}\:\text{s}^{-2}$, $\gamma = 10^{-17}\:\text{kg}\;\text{s}^{-1}$, and $\eta = 22\;\text{kg}\;\text{m}^{-2}\:\text{s}^{-1}$. At low drive power, $V_{g,ac} = 50\; \mu\text{V}$, the time constant is $72\; \mu\text{s}$. At higher drive power, $V_{g,ac} = 500\; \mu\text{V}$, the resonator non-linearity yields a more rapid response time of $16\; \mu\text{s}$.}
\label{fig4}
\end{figure}

To realize fast response in sensing applications, the CNT resonator can be driven into the nonlinear regime to respond faster than the linear time constant $\tau = Q/\omega_0$. Figure \ref{fig4} shows simulation results to illustrate this point, using device parameters from figure \ref{fig2}. The spring constant, $k$, is instantaneously shifted at $t = 0$ by a $\Delta k$ so that resonance frequency, $\omega_0 = \sqrt{k/m}$, is decreased by $\Delta \omega_0=2\pi\times 10\; \text{kHz}$.

This simulates a shift of the mechanical resonance frequency induced by a change in the CNT tension. This could be generated, for example, by the spin flip of a single molecule magnet in a magnetic field gradient \cite{Willick2014}. In both panels, for $t < 0$, the device is driven at a frequency slightly below the mechanical resonance, so that the amplitude of motion is below the maximum value. The change in resonance frequency at $t = 0$ brings the CNT closer to resonance and increases the amplitude of motion, in this case by $\Delta x_0 \approx 0.3\;\text{nm}$. This change in amplitude would change the measured current by $\sim 0.2\;\text{nA}$ for a similar conductance peak as the one shown in figure \ref{fig2}. The top panel of figure \ref{fig4} shows the response for a weak drive in which the device is dominated by linear terms, and the lower panel shows the case for a stronger drive in which the non-linear terms dominate. The weakly driven linear response time constant is $72\;\mu\text{s}$, whereas under strong driving it is $16\;\mu\text{s}$, about 5 times faster. 

In conclusion, fast electrical readout of the mechanical motion of a suspended CNT resonator is enabled by a proximate HBT circuit operated as a linear AC current amplifier. A Coulomb rectification measurement made it possible to explore the strongly non-linear regime. The transient mechanical response was observed under pulsed driving, including fast turn-on response in the non-linear regime. The resonator behaviour is well described by Duffing and non-linear damping terms in equation \ref{eq:ho}. For the circuit and CNT device presented, a response time $\sim 10\; \mu\text{s}$ to an instantaneously applied force could be obtained under strong driving. As an example, this could permit single-shot readout of the magnetic states of single magnetic molecules on timescales comparable to their spin relaxation and/or dephasing times. 



\section*{Supplementary Material}

The supplementary material shows the methods used to extract the time-domain RMS envelope and to fit this data to the non-linear harmonic oscillator equation, as well as conductance data from the Coulomb peak measured in figure \ref{fig3}.

\acknowledgments{We thank Greg Holloway, Andrew Ward, and Gaganprit Gill for assistance. This work was supported by NSERC and the Ontario Ministry for Research \& Innovation. This work benefitted from the University of Waterloo's Quantum NanoFab, supported by the Canada Foundation for Innovation, Industry Canada, and Mike \& Ophelia Lazaridis.}

\clearpage

\renewcommand{\theequation}{S\arabic{equation}}
\setcounter{equation}{0}  
\renewcommand{\thefigure}{S\arabic{figure}}
\setcounter{figure}{0}
\begin{widetext}
\section*{SUPPLEMENTARY INFORMATION: Fast measurement of carbon nanotube resonator amplitude with a heterojunction bipolar transistor}

\subsection*{$I_{out}$ envelope extraction}

The time-domain measurements reported in figure 2(c,d) and figure 3 were measured with an oscilloscope after amplification of $I_{out}$. The current recorded at the oscilloscope is a product of the conductance of the CNT and the 100 kHz bias voltage. To extract the conductance of the CNT, we apply an RMS envelope extraction with $10\mu\text{s}$ window, analogous to the output signal of a lock-in amplifier with similar time constant. 

An example of the oscilloscope signal is shown in figure \ref{figRMS}a. In addition to the 100 kHz bias frequency, there is a 16 kHz oscillation in the average current that was observed in all measurements and is likely an artifact of the voltage source used for $V_{g,dc}$. This noise is removed during post-processing by applying a narrow band-stop filter at 16 kHz. Figure \ref{figRMS}b shows the data from (a) after this filtering. Finally, the RMS envelope is extracted from the filtered data. Figure \ref{figRMS}c shows the envelope extracted from (b).

\begin{figure}[h]
\includegraphics[width=0.51\linewidth]{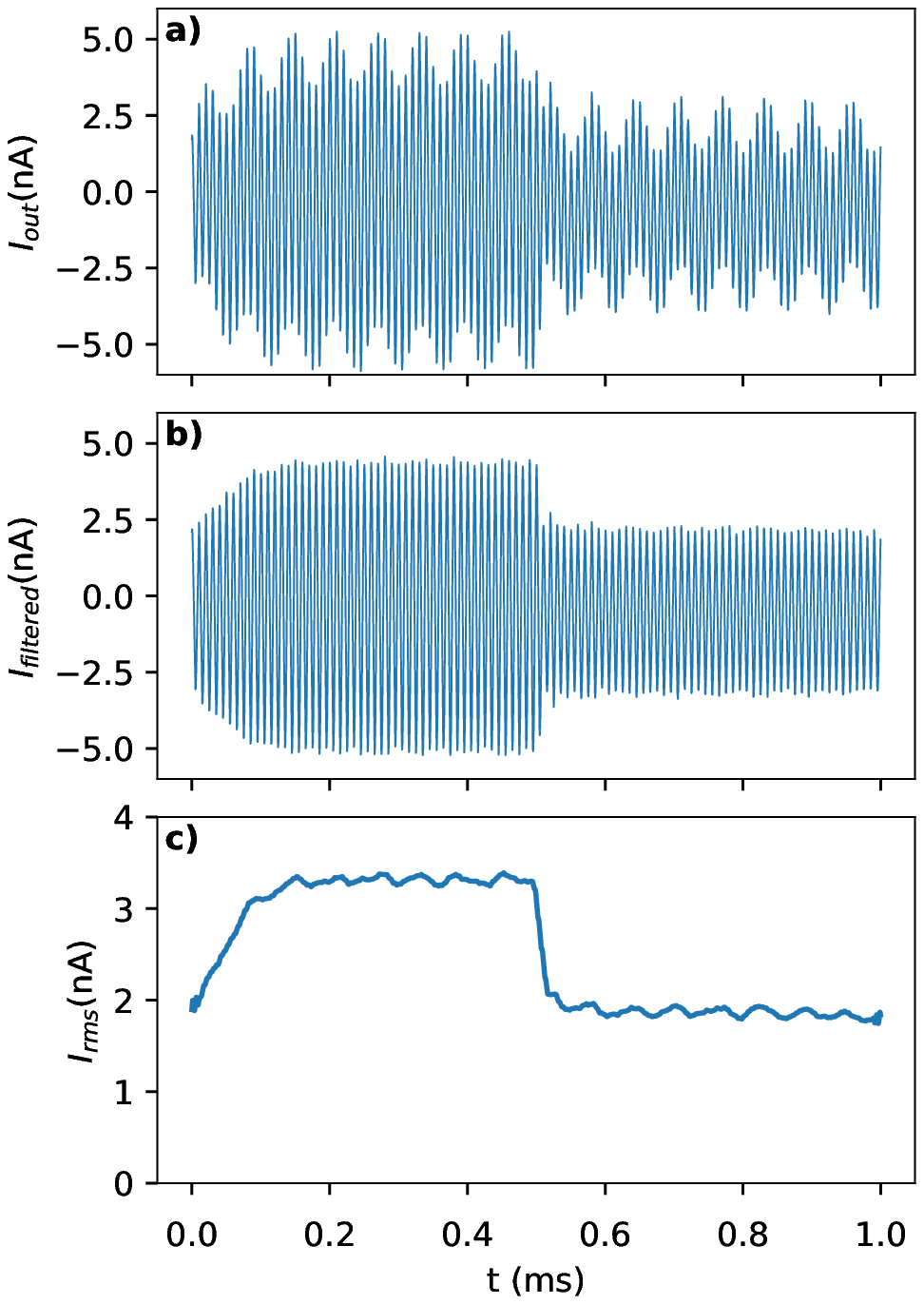}
\caption{\textbf{a} $I_{out}$ as recorded on an oscilloscope and averaged over 2048 shots. \textbf{b} The same data as (a) after applying a band-stop filter at 16 kHz. \textbf{c} The RMS envelope of (b) extracted using a $10 \mu\text{s}$ sliding window.}
\label{figRMS}
\end{figure}

\subsection*{Fitting time-domain data}

To model the time-domain current measurements presented in figures 2 and 3 of the main text, we use equation (2) to determine the oscillator motion as a function of time, then calculate current using equation (1). The parameters used in these equations are determined as follows:

\begin{itemize}
\item $C_g$, $\partial C_g/\partial x$, and $m$ are estimated from the device geometry
\item From off-resonant current data, e.g. at $f = 87\;\text{MHz}$ in figure 2a of main text, $\partial^2 I' / \partial V_g^2$ is calculated.
\item The resonant frequency, $\omega_0$, is determined using low drive power and is used to calculate $k$.
\item Values for $\alpha$, $\gamma$, and $\eta$ are estimated from similar devices in literature, then fine tuned to fit the time domain data in the following sequence:
\begin{enumerate}
\item The transient behaviour when $F_{drive}$ is turned off is fit by adjusting damping parameters $\gamma$ and $\eta$.
\item The fit to turn on behaviour is improved by adjusting $\alpha$.
\item $\gamma$ and $\eta$ are adjusted to fit the steady state amplitude while maintaining a good match to the turn-off behaviour. 
\end{enumerate}
\end{itemize}

Figure \ref{figFit} shows the fitting steps for the data from main text figure 2d. 

\begin{figure}
\includegraphics[width=0.51\linewidth]{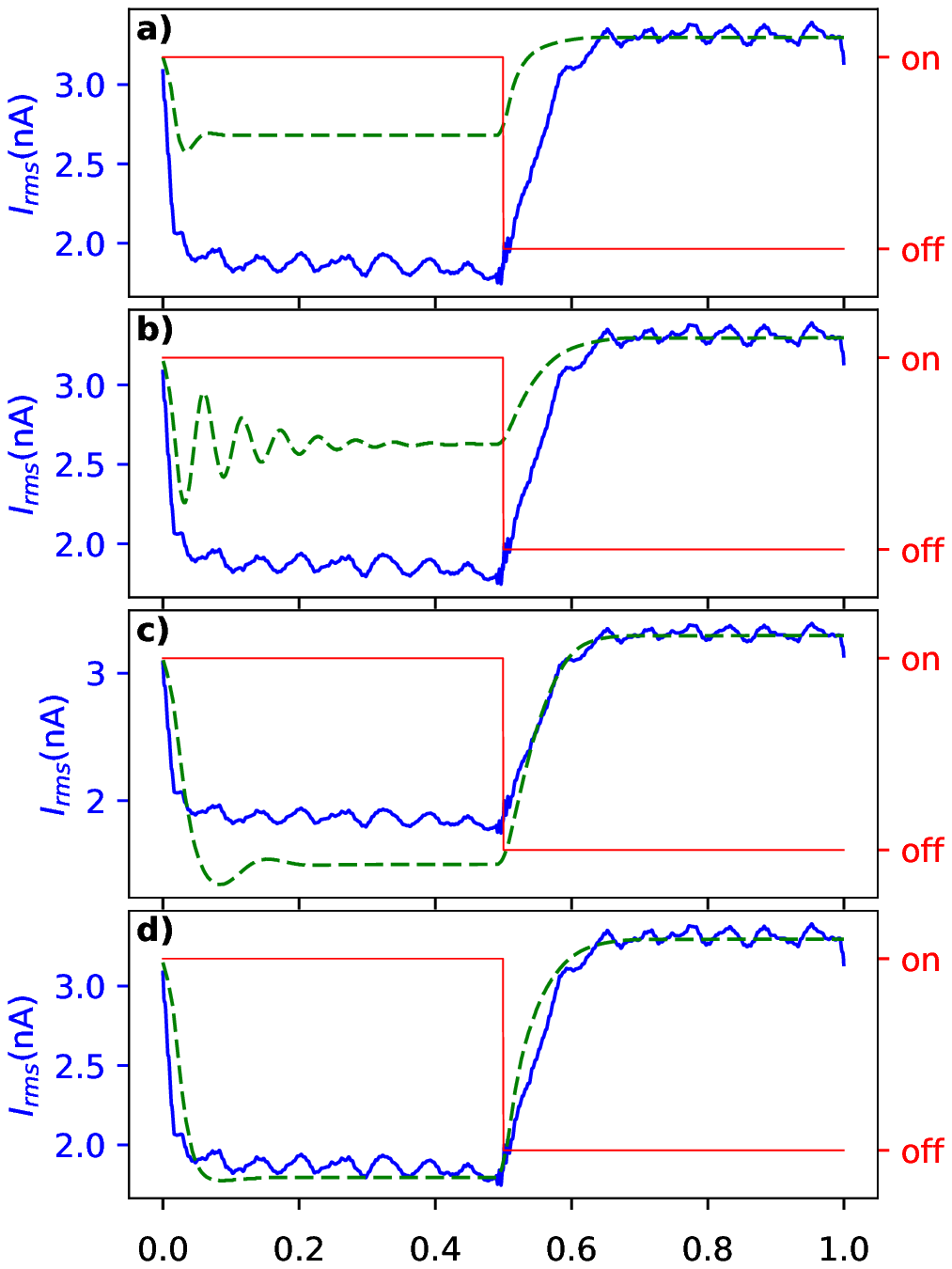}
\caption{Fitting of time-domain current signals. \textbf{a} Using rough initial estimates of $\alpha = 5 \times 10^{10}, \gamma = 1 \times 10^{-16}, \eta = 100$. \textbf{b} After fitting the turn-off transient, $\gamma = 1 \times 10^{-16}, \eta = 4$. \textbf{c} Fitting the turn-on transient, $\alpha = 5 \times 10^9$. \textbf{d} To adjust how damping affects turn-on and steady state, while maintaining the turn-off fit, $\gamma = 1 \times 10^{-17}, \eta = 22$}
\label{figFit}
\end{figure}

\subsection*{Coulomb peak used for figure 3}

Figure \ref{figCond} shows the conductance and mechanical resonance signal across the Coulomb peak used for figure 3 of the main text.

\begin{figure}
\includegraphics[width=0.51\linewidth]{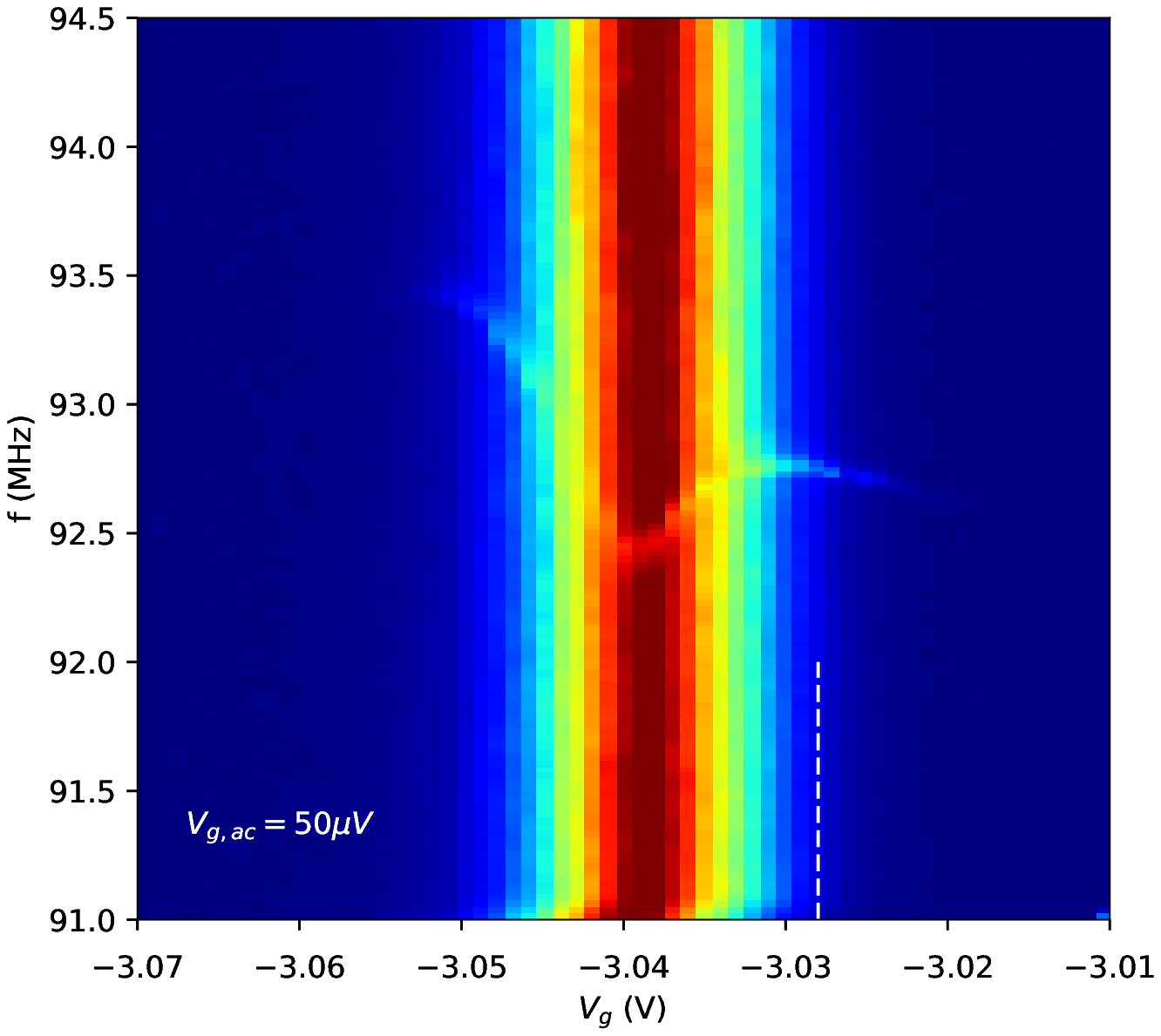}
\caption{Current through the CNT mechanical resonator versus DC gate voltage and driving frequency. The color scale represents $I_{out}$ from 0 (dark blue) to 10 nA (red). The AC driving signal is $V_{g,ac} = 50\;\mu\text{V}$. The white dashed line indicates the gate voltage at which figure 3 of the main text is measured.}
\label{figCond}
\end{figure}

\pagebreak
\end{widetext}

\clearpage

\end{document}